\documentclass[11pt,a4paper]{article}
\usepackage[margin=2cm]{geometry}
\usepackage[utf8]{inputenc}
\usepackage{amsmath}
\usepackage{amsthm}
\usepackage{amssymb}
\usepackage{placeins}
\usepackage{indentfirst}
\usepackage{microtype}
\usepackage{graphicx}
\usepackage{subcaption}
\usepackage{setspace}
\usepackage[
breaklinks=true,colorlinks=true,
linkcolor=blue,urlcolor=blue,citecolor=blue,
bookmarks=true,bookmarksopenlevel=2]{hyperref}

\title{Construction of symplectic systems from parameter-drift Hamiltonian maps}
\author{Gabriel C. Grime$^{1,2}$ \and Philip J. Morrison$^{1}$}
\date{\small
$^1$Department of Physics and Institute for Fusion Studies,\\
The University of Texas at Austin, Austin, TX, 78712, USA\\
$^2$ Institue of Physics, University of São Paulo, São Paulo, SP, 05508-220, Brazil\\  
\today}

\begin{document}

\maketitle

\begin{abstract}
    \noindent We reveal the symplectic nature of  parameter-drift  maps by embedding them into extended phase space. Applying the embedding to the parameter-drift standard nontwist map, our construction yields an autonomous symplectic map in extended phase space that preserves key dynamics of the original system. Computing finite time Lyapunov exponents, the symplectic map shows limitations of ensemble-based diagnostics, common in the parameter-drift literature, and provides new insights into transport phenomena in these nonautonomous systems.
\end{abstract}

\section{Preliminaries and motivation}

The dynamics of deterministic systems under parameter drift have attracted increasing attention, especially for modeling time-dependent phenomena such as climate change \cite{pierini2018} and magnetically confined plasmas \cite{janosi2024magnetic}. In particular, low-dimensional Hamiltonian-like systems have been studied under slow parameter variation \cite{janosi2019,janosi2021,janosi2022,janosi2024,marcos2025}.

To capture the nonautonomous nature of these systems, the Ensemble-Average Pairwise Distance (EAPD) was introduced as a measure of chaos strength in ensembles of orbits at fixed times—so-called snapshot objects \cite{janosi2019}. EAPD distinguishes between regular and chaotic behavior and has been used to detect the breakup of snapshot tori. However, this ensemble-based approach has key limitations: it obscures individual orbit dynamics that may contradict ensemble trends, potentially leading to misleading conclusions about torus breakup. Furthermore, time-dependent Lyapunov exponents, defined from the time derivative of the EAPD, may fail to accurately represent the behavior of the majority of orbits in the ensemble.

In this work, we address these limitations by constructing symplectic maps that reveal the autonomous structure of parameter-drift Hamiltonian systems. Using a suitable generating function, we embed the nonautonomous map into the extended phase space, increasing its dimensionality while demonstrating the symplectic character of the systems. This framework allows for the proper computation of Lyapunov exponents for each orbit within an ensemble. Moreover, by comparing finite-time Lyapunov exponents with those derived from the EAPD method, we assess how well ensemble-based diagnostics capture individual orbit behavior.

We consider a parameter-drift version of the Standard Nontwist Map (SNM) \cite{diego1993},%
\begin{subequations}\label{eq:snm}
    \begin{align}
        x_{n+1} &= x_n+ a(1-y_{n+1}^2) \mod{1}\\[0.2cm]
        y_{n+1} &= y_n - b\sin{(2\pi x_n)},
    \end{align}
\end{subequations}
a two-dimensional symplectic map with parameters $a$ and $b$ and phase space coordinates $(x,y)\in\mathbb{S}^1\times\mathbb{R}$. The symplectic structure is ensured by the existence of a generating function. For instance, the SNM has  a generating function of the second-kind  given by
\begin{equation}\label{eq:genfuncsnm}
    F(x_n,y_{n+1}) = y_{n+1}x_n+ ay_{n+1}\left(1 - \dfrac{y_{n+1}^2}{3}\right) - \dfrac{b}{2\pi}\cos{(2\pi x_n)},
\end{equation}
where $(x_n,y_n)$ stand for the ``old'' and $(x_{n+1},y_{n+1})$ for the ``new'' canonical coordinate and momentum. The resulting canonical relations are
\begin{subequations}
    \begin{align}
        y_n &= \dfrac{\partial F}{\partial x_n} = y_{n+1} + b\sin{(2\pi x_n)}\\[0.1cm]
        x_{n+1} &= \dfrac{\partial F}{\partial y_{n+1}} = x_n+ a - ay_{n+1}^2,
    \end{align}
\end{subequations}
which recover Eq.~\eqref{eq:snm}. Introducing parameter drift by allowing $b$ to be a function of the iteration integer $n$, we obtain the time-dependent two-dimensional map
\begin{subequations}\label{eq:2d_map_drift}
    \begin{align}
        x_{n+1} &= x_n + a(1-y_{n+1}^2) \mod{1}\\[0.2cm]
        y_{n+1} &= y_n - b(n)\sin{(2\pi x_n)},
    \end{align}
\end{subequations}
where $n_0<n<n_\mathrm{max}$.  {For simplicity, we will use  linear drift, $b(n)=b_0+\beta n$.}

\begin{figure}[htb]
    \centering
    \includegraphics[width=0.99\textwidth]{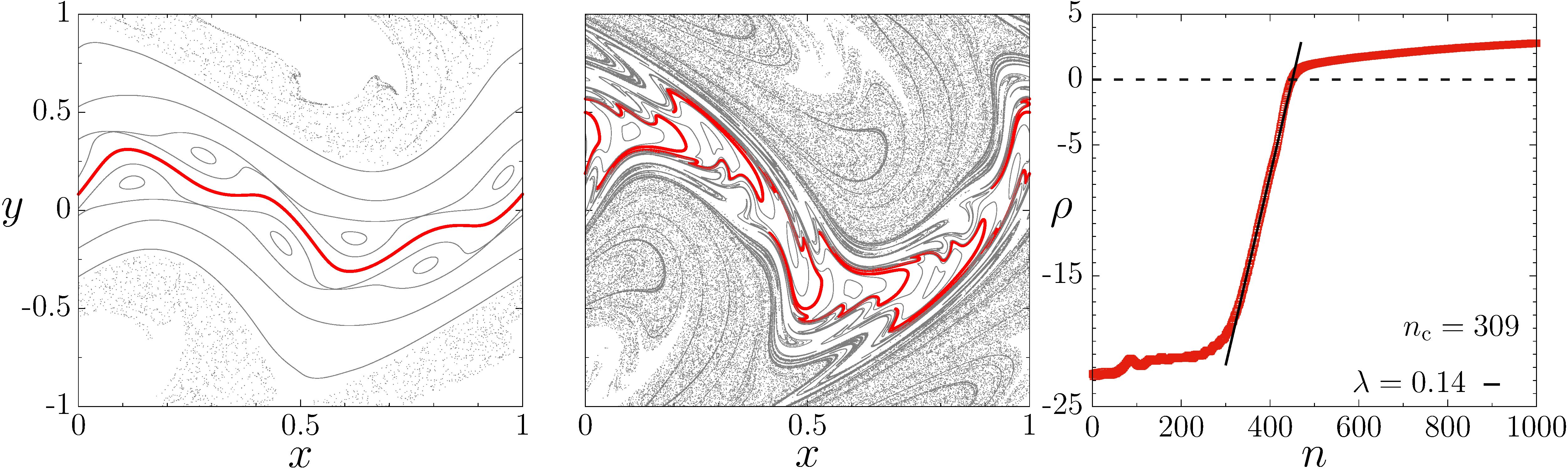}
    \caption{Snapshots of the non-autonomous SNM corresponding to (left) $n=20$ and (center) $n=70$. The parameters are $a=0.345$, $b_0=0.3$ and $\beta=0.007$. (Right) Ensemble average pairwise distance ($\rho$) for the same $a$ and $b_0$, with $\beta=0.0015$.}
    \label{fig:2d_param_drift}
\end{figure}

Previous works have investigated nontwist Hamiltonian maps subjected to parameter drift \cite{janosi2024chaos}, including the standard nontwist map \cite{marcos2025}. Preparing an ensemble of initial conditions located on an invariant torus of the frozen system ($b=b_0$),  it is observed that under parameter drift the tori in the nonautonomous maps drift across the phase space, as shown in Figure~\ref{fig:2d_param_drift}(a) and \ref{fig:2d_param_drift}(b). As the perturbation parameter increases, the snapshot tori eventually break up, changing their topology.

To quantify the degree of chaos in parameter-drift systems, the Ensemble-Average Pairwise Distance (EAPD) has been introduced~\cite{janosi2024}. It is defined as
\begin{equation}
    \rho=\langle \ln{r_n}\rangle
\end{equation}
where the average $\langle\cdot\rangle$ is taken over an ensemble of trajectories for which $r_n$ stands for the distance between neighboring trajectories within the ensemble at discrete times $n$. The evolution of $\rho(n)$ is illustrated in Figure \ref{fig:2d_param_drift}(c) for the snapshot shearless torus. Using this measure, a global threshold for the breakup of snapshot tori is defined at the transition between the initial plateau and the linear regime, at $n_\mathrm{c}\approx 309$. This marks the onset of exponential orbit separation, on average, within the ensemble. The slope in this regime is interpreted as the instantaneous Lyapunov exponent, in this case, $\lambda=0.14$.

\section{Results}

 {It was not pointed out in  previous works that  parameter drift Hamiltonian systems still possess  symplectic structure. Taking the parameter-drift standard nontwist map as an example, we define an iteration-dependent generating function, equivalent to a time-dependent generating function in continuous systems, given by
\begin{equation}\label{eq:GFn}
    F(x_n,y_{n+1};n) = F(x_n,y_{n+1};b(n)) = y_{n+1}x_n+ ay_{n+1}\left(1 - \dfrac{y_{n+1}^2}{3}\right) - \dfrac{b(n)}{2\pi}\cos{(2\pi x_n)},
\end{equation}
which yields the map \eqref{eq:2d_map_drift}. Therefore, in general, all parameter-drift Hamiltonian maps still preserve their symplecticity; however, they are not autonomous. In order to recover time independence, we resort to the extended phase space, as detailed below.
}

{To construct a four-dimensional symplectic map derived from the parameter-drift standard nontwist map, we start from the iteration-dependent generating function \eqref{eq:GFn}, identifying the iteration as a new coordinate $q$.} Let $(x,q,y,p)$ be canonical coordinates of a four-dimensional Hamiltonian map where $(y,p$) are  the canonical momenta conjugate  to the coordinates $(x,q$). The simplest generating function is an extension of Eq.~\eqref{eq:GFn}, given by
\begin{equation}\label{eq:4dgenfunc}
    G(x_n,q_n,y_{n+1},p_{n+1}) = F\left(x_n,y_n;b(q_n)\right) + (q_n+1)p_{n+1}.
\end{equation}
Here, the first term refers to the generating function \eqref{eq:genfuncsnm}, considering the perturbation parameter a function of the coordinate $q$, interpreted as a time coordinate. The last term in Eq.~\eqref{eq:4dgenfunc} is based on the identity generating function but has the purpose of adding one unit to $q$ on each iteration.

This generating function relates old and new variables via
\begin{align*}
    y_n &= \dfrac{\partial F}{\partial x_n} = y_{n+1} + b(q_n)\sin{(2\pi x_n)}\\[0.1cm]
    p_n &= \dfrac{\partial F}{\partial q_n} = -\dfrac{b'(q_n)}{2\pi}\cos{(2\pi x_n)} + p_{n+1}\\[0.1cm]
    x_{n+1} &= \dfrac{\partial F}{\partial y_{n+1}} = x_n+ a(1 - y_{n+1}^2)\\[0.1cm]
    q_{n+1} &= \dfrac{\partial F}{\partial p_{n+1}} = q_n + 1
\end{align*}
Rewriting these equations, we obtain the four-dimensional nontwist map
\begin{equation}\label{eq:4D_general}
    \begin{cases}
        x_{n+1} &= x_n+ a(1-y_{n+1}^2)\\
        q_{n+1} &= q_n + 1\\
        y_{n+1} &= y_n - b(q_n)\sin{(2\pi x_n)}\\
        p_{n+1} &= p_n + \dfrac{b'(q_n)}{2\pi}\cos{(2\pi x_n)}
    \end{cases}
\end{equation}

{We have shown that parameter drift symplectic maps, when lifted to  the extended phase space, produce autonomous symplectic systems. With similar and straightforward arguments, the same  conclusion follows  for Hamiltonian flows, like the parameter-drift Hamiltonian Duffing oscillator \cite{janosi2019}. Let $H(p_1,\dots,p_m,q_1,\dots,q_m,\mu(t),t)$ be a time-dependent parameter-drift Hamiltonian function. In extended phase space, where we identify $q_{m+1}\equiv t$ and $p_{m+1}\equiv -H$, the new Hamiltonian $\mathcal{H}(p,q)=p_{m+1}+H(p_1,\dots,p_m,q_1,\dots,q_m,\mu(q_{m+1}),q_{m+1})$ yields autonomous canonical equations of motion.}

In parameter drift systems, the Ensemble-Average Pairwise Distance (EAPD) and its associated growth rate $\lambda$, do not provide a true measurement of the Lyapunov exponent. However, since our four-dimensional map is autonomous, we can compute the full Lyapunov spectrum of individual orbits directly using Wolf's method \cite{wolf1985}, as well as the Finite-Time-Lyapunov-Exponents (FTLE). This allows a meaningful comparison between the orbit-wise FTLE and the ensemble-average growth rate.

\begin{figure}[htb]
    \centering
    \includegraphics[width=0.99\linewidth]{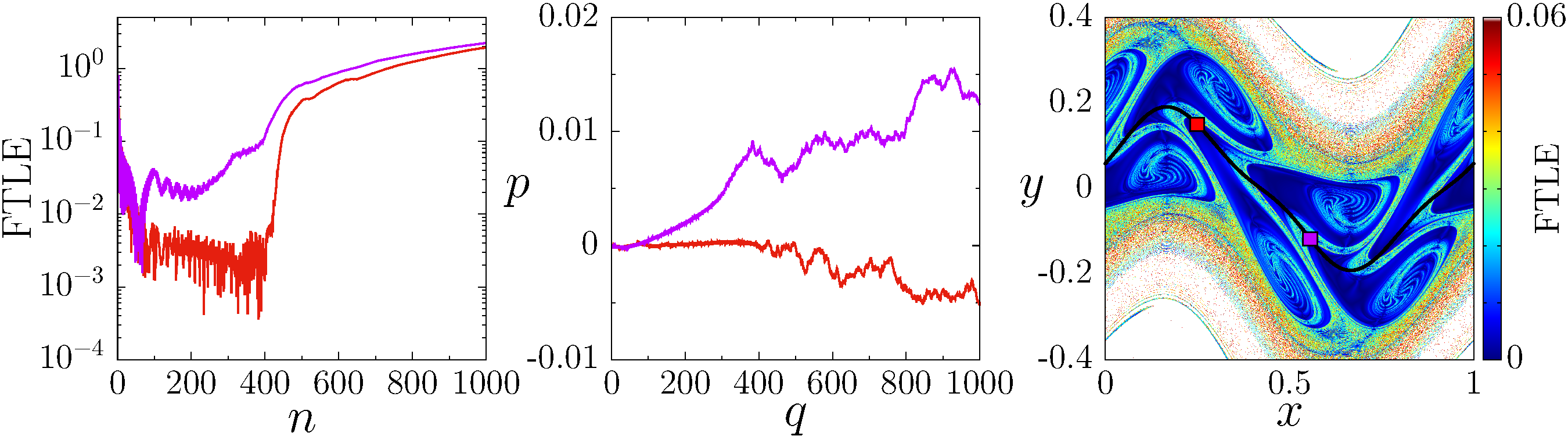}
    \caption{\label{fig:2}(left) Finite Time Lyapunov Exponent
    for two different initial conditions on the same ensemble. (center) $(q,p)$ section of phase space. (right) FTLE in the $(x,y)$ plane. Parameters are $a=0.345$, $b_0=0.3$, $b_\mathrm{max}=1.8$, $n_\mathrm{max}=1000$, $n=30$.}
\end{figure}

Unlike the EAPD, which captures ensemble behavior, the FTLE quantifies the chaos strength of each orbit at each step. Figure~\ref{fig:2}(a) shows the evolution of the maximum FTLE as a function of the iteration $n$ for two different initial conditions located on the shearless torus. Although both start on the same snapshot torus, their FTLE evolves differently. They have distinct stickiness strengths, and their transition to chaos occurs at different times.

These differences within the same ensemble arise from the presence of coherent structures in phase space. By computing the FTLE over a grid of initial conditions in the $(x,y)$ plane, the resulting Lagrangian Coherent Structures (LCS) are revealed, as shown in Figure~\ref{fig:2}(c). Despite belonging to the same ensemble (plotted in black), the red and magenta orbits lie on different coherent structures and thus exhibit distinct FTLE behavior.

In addition to the $x$ and $y$ coordinates, the autonomous map includes a new canonical pair, $(q,p)$, whose section is shown in Figure~\ref{fig:2}(b) for the same orbits, which can provide further insight into the evolution of these orbits in phase space.

\begin{figure}[htb]
    \centering
    \includegraphics[width=0.99\linewidth]{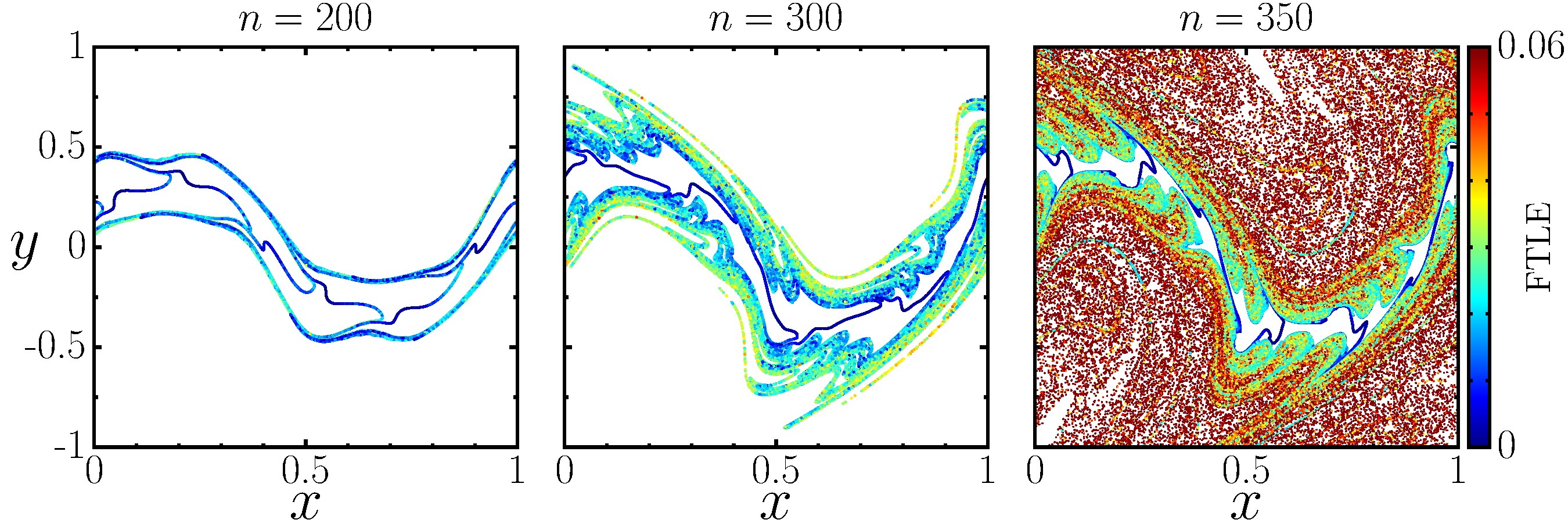}
    \caption{\label{fig:3}Snapshot of parameter-drift standard nontwist map after $n$ iterations. Colors represent the maximum FTLE after $n$ iterations.}
\end{figure}

To examine how the FTLE is distributed over time across the ensemble, Figure~\ref{fig:3} presents snapshots of the ensemble orbits at various iteration steps \(n\). Although the ensemble as a whole initially retains some features of a torus-like topology, subsets of initial conditions begin to exhibit chaos. While some orbits still display near-zero FTLE (dark blue points) and maintain a torus-like structure, others exhibit chaos, characterized by larger FTLE (cyan and warmer colors).

This coexistence of regular and chaotic populations highlights the diverse dynamical responses within the same ensemble. Moreover, the critical iteration for torus breakup predicted by the EAPD, \(n_\mathrm{c} = 309\), does not fully capture the complexity of the transition from regular to chaotic motion on an orbit-by-orbit basis. Even when orbits belong to the same ensemble, they are distributed in populations whose dynamical evolution can differ significantly.

\begin{figure}[htb]
    \centering
    \includegraphics[width=0.99\linewidth]{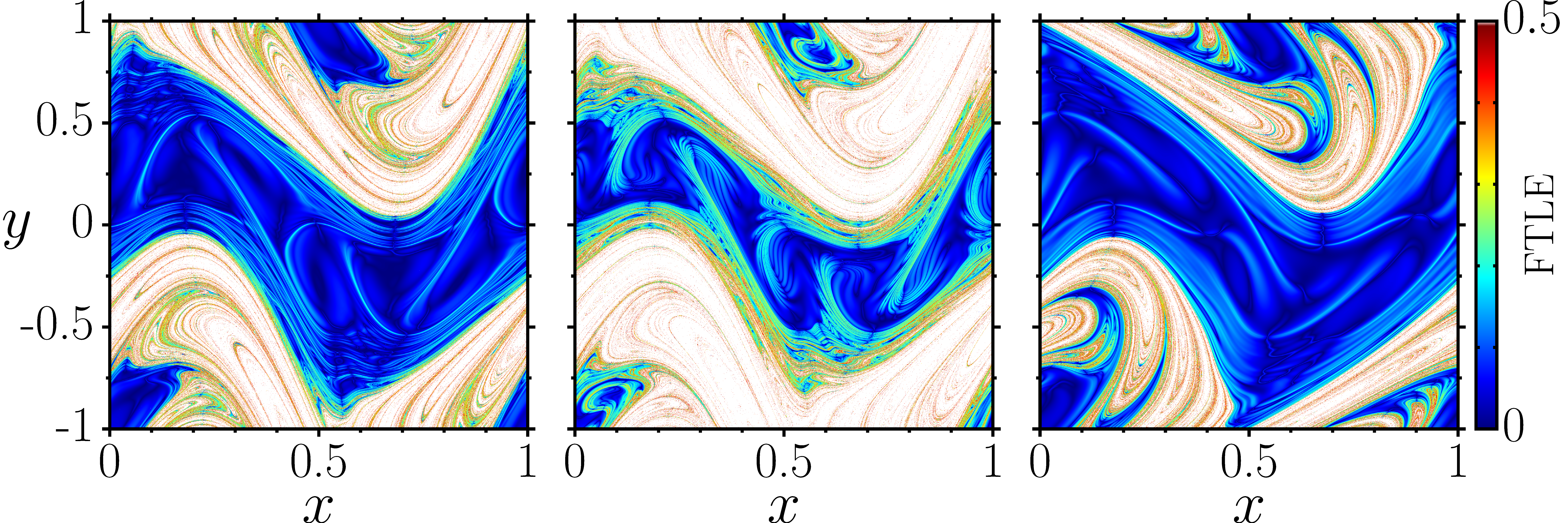}
    \caption{\label{fig:LCS}(left) Repelling LCS in frozen map. Repelling (center) and attracting (right) LCS in four-dimensional map. $a=0.36$, $b_0=0.6$, $\beta=0.008$ and $n=30$.}
\end{figure}

To further investigate the underlying mechanisms responsible for the heterogeneous behavior of orbits within the same ensemble, we analyze the Lagrangian Coherent Structures (LCS) for a different set of parameters, in Figure \ref{fig:LCS}. These structures provide a geometric interpretation of transport and mixing in the phase space, highlighting regions of stretching and folding that characterize chaotic dynamics \cite{haller2015}. In particular, LCS offers a complementary view to the snapshot torus deformation, since these structures have the same role as time-dependent foliations obtained in parameter-drift systems \cite{janosi2024}.

The intricate repelling LCS, shown in Figure \ref{fig:LCS}(b), might be related to the distortions in the ensembles of trajectories occurring in parameter-drift systems, like the red torus distortion in Fig.~\ref{fig:2d_param_drift}. Complementary, the attracting LCS are shown in Figure \ref{fig:LCS}(c).

\FloatBarrier
\section{Conclusions}

The results here indicate that the ensemble average approach in Hamiltonian parameter-drift systems can only provide typical information about ensembles of trajectories in phase space.

We have introduced a four-dimensional autonomous symplectic map derived from the parameter-drift standard nontwist map by promoting the time to a canonical coordinate. This construction reveals the Hamiltonian structure underlying parameter-drift systems and enables the application of traditional chaos quantifiers, such as Lyapunov exponents, in the new autonomous dynamics. By comparing the Finite-Time Lyapunov Exponents (FTLE) of individual orbits with the Ensemble-Average Pairwise Distance (EAPD) growth rate, we demonstrated that the EAPD captures global trends but fails to resolve transitions between regular and chaotic behavior. It also might lead to misleading results, since chaotic orbits hide the regularity of significant populations of orbits in the ensemble.

The spatial organization of FTLE across the ensemble revealed the presence of Lagrangian Coherent Structures (LCS). These structures expose the geometric features underlying phase space transport and clarify why orbits in the same ensemble can exhibit different dynamics. In particular, the complex LCS in the autonomous map provide insight into the mechanisms by which parameter drift induces deformation of snapshot tori. Finally, this framework can open a path for deeper understanding of parameter-drift Hamiltonian systems through the lens of symplectic structures.

\bibliographystyle{unsrt}

\end{document}